\documentstyle[aps,preprint,psfig,epsf]{revtex}

\input epsf

\tightenlines

\begin{document}

\title{\Large{Structure Functions and Pair Correlations of the 
Quark-Gluon Plasma}}

\author{Markus H. Thoma}

\address{Max-Planck-Institut f\"ur extraterrestrische Physik,
P.O. Box 1312, 85741 Garching, Germany}

\maketitle

\vspace{0.4in}

\begin{abstract}
Recent experiments at RHIC and theoretical considerations indicate that 
the quark-gluon plasma, present in the fireball of relativistic heavy-ion
collisions, might be in a liquid phase. The liquid state can be identified
by characteristic correlation and structure functions. Here definitions 
of the structure functions and pair correlations 
of the quark-gluon plasma are presented as well 
as perturbative results. These definitions might be useful for verifying
the quark-gluon-plasma liquid in QCD lattice calculations.  
\end{abstract}

\vspace{0.2in}

\section{Introduction}

Relativistic heavy-ion collision experiments at RHIC have found evidences for
a new state of matter, the quark-gluon plasma (QGP) \cite{QM,Phenix}. It is 
expected to
exist in the hot phase of the fireball in ultrarelativistic nucleus-nucleus collisions
as well as in the early Universe for the first few microseconds after the Big
Bang at temperatures above $T_c=170$ MeV \cite{Karsch1}. In the limit of an 
infinitely
high temperature the QGP should be an ideal gas because the effective
temperature dependent coupling constant becomes small due to asymptotic 
freedom. Therefore in the (extremely) high temperature regime the interaction 
between quarks and gluons is weak and the QGP can be described by perturbative
methods \cite{Thoma}. However, at temperatures which can be realized in 
accelerator experiments, i.e., at maximum a few times of the transition 
temperature, the QGP is a strongly interacting many-body system. 

In classical, non-relativistic electromagnetic plasmas one distinguishes between
weakly coupled and strongly coupled plasmas. For this purpose the Coulomb coupling 
parameter, $\Gamma = Q^2/(dT)$, where $Q$ is the charge of the plasma particles, 
$d$ the inter-particle distance, and $T$ the plasma temperature, 
is considered \cite{Ichimaru}. In the case of an one-component plasma with
a pure Coulomb interaction the Coulomb coupling parameter corresponds to the
ratio of the interaction energy to the thermal energy per particle.
Most plasmas in nature and in the laboratory 
are weakly coupled, i.e., $\Gamma \ll 1$. If $\Gamma $ is not much smaller than one,
the plasma is called strongly coupled. These plasmas can be in a gas, liquid, or
even solid phase \cite{Ichimaru}. For $\Gamma > {\rm O}(1)$ the plasma 
behaves like a liquid and for $\Gamma > 172$ the plasma particles are predicted to
arrange in ordered structures, the plasma crystal \cite{Ichimaru}. The latter state was 
discovered in so-called complex or dusty plasmas \cite{Thomas}.

In real plasmas the Coulomb interaction is modified to a Yukawa type interaction due 
to screening. Then the Coulomb coupling parameter, defined above, is no longer the 
ratio of the interaction to thermal energy. An additional parameter, the ratio
of the inter-particle distance to the Debye screening length, $\kappa = d/\lambda_D$,
becomes important. If $\kappa \gg 1$ the plasma is weakly coupled. Now the 
transitions to the liquid and solid phases are considered within a phase diagram in the 
$\Gamma$-$\kappa$-plane (see e.g. Ref.\cite{Hamaguchi}). For larger values of
$\kappa $ a higher value of $\Gamma $ is required to achieve a phase transition to a more
ordered phase (gas-liquid or liquid-solid).
However, as long as $\kappa$ is of the order of one or smaller,
the phase transitions are not much shifted to higher values of $\Gamma $ \cite{Hamaguchi}.
In addition, it should be noted that in strongly coupled plasmas the screening potential
contains a long-range power law contribution \cite{Ichimaru}. 
 
In the case of a QGP the coupling parameter is estimated to be 
$\Gamma = 2Cg^2/(4\pi dT)=1.5$ - 5 \cite{Thoma1,Thoma2}. Here $C$ 
is the Casimir invariant ($C=4/3$ for quarks and $C=3$ for gluons), $d\simeq 0.5$ fm
the inter-particle distance, and $T$ the temperature assumed to be about 200 Mev 
corresponding to a strong coupling constant $g\simeq 2$. The factor 2 in the
numerator comes from taking into account the magnetic interaction in addition 
to the static electric (Coulomb) interaction, which are of the same magnitude
in ultrarelativistic plasmas.  

As a matter of fact, the true coupling parameter of the QGP might be even larger because
the potential might differ from a simple Coulomb or Yukawa potential corresponding
to a one-gluon exchange. Indeed the cross section enhancement by about a factor of 80 
due to non-perturbative effects discussed in Ref.\cite{Molnar} indicates a stronger 
interaction potential. The fact that the cross section enhancement by a factor 2 - 9
due to modifications of the scattering for $\Gamma =$1.5 - 5 \cite{Thoma2} does not
explain the observed cross section enhancement suggests that a realistic coupling
parameter could be up to an order of magnitude larger.  

Following the investigations on non-relativistic plasmas,
we proposed that the QGP in relativistic
heavy-ion collisions is a liquid rather than a gas \cite{Thoma2} which
is supported by experimental results \cite{Phenix} as well as other
theoretical considerations \cite{Shuryak,Gyulassy,Heinz,Cassing}. Among those are an 
observed  strong elliptic flow, a strong jet quenching and a fast thermalization 
corresponding to large parton cross sections, and a small viscosity
indicating the QGP as an ideal fluid. 

The presence of a liquid QGP expands the phase diagram of strongly
interacting matter by a new phase transition from a QGP liquid to a gas
at high temperatures. Also a new critical point in the phase 
diagram should show up, above which only a supercritical fluid exists. Assuming that
this phase transition occurs at $\Gamma  \simeq 1$ - the distance parameter in the QGP, 
$\kappa = 1$ - 3, is sufficiently small - and that the temperature
dependent coupling constant decreases logarithmically with the temperature, 
the transition temperature is estimated to be of the order of a few $T_c$. 
Such a temperature could be reached at LHC, where the liquid-gas transition might occur
during the expansion of the fireball \cite{Peshier}. 

The phase diagram of strongly interacting matter can also be studied within
lattice QCD. However, the phase transition from the liquid to the gas phase
at high temperatures is probably difficult to observe in lattice calculations
\cite{Karsch2}. However, correlation functions, which can be investigated
by lattice simulations, can also provide information on the state of the QGP.     

\section{Correlation functions in liquids}

Quantitative investigations of liquids are possible by considering
correlation functions \cite{Hansen}. 
In particular, the pair correlation or radial
distribution functions in coordinate space as well as the dynamic and static
structure functions in momentum space show characteristic properties
within the different phases. For example, in the case of a liquid the pair 
correlation and its Fourier transform, the static structure function, 
exhibit a pronounced peak and one or two small and broad additional peaks. The first
peak corresponds in coordinate space to the inter-particle distance, which 
is fixed in an incompressible liquid corresponding to a short-range order. 
In the case of a solid crystalline phase, where also a long-range order
exists, a number of sharp peaks are observed. In the gas phase, where
no order is present, the pair correlation function shows no clear structures.

Here we want to consider the static structure function, which
shows a similar peak structure as the pair correlation function \cite{Hansen}. Furthermore we
will also discuss the dynamic structure function giving additional 
information about the system. First we will define these structure functions.
These definitions might be in particular useful for confirming and
investigating the liquid 
phase of the QGP in QCD lattice simulations. To demonstrate the
structure functions and for comparison with future lattice calculations 
we will calculate them perturbatively. Of course, in this case we do not 
expect any liquid behavior, e.g., a single clear peak, as the perturbative
regime corresponds to the gas phase of the QGP with $\Gamma \ll 1$.

The static density-density autocorrelation function is defined as \cite{Hansen,Ichimaru2}
\begin{equation}
G({\bf r}) = \frac{1}{N}\> \int d^3r' \> \langle \rho ({\bf r} + {\bf r'},t)
\rho ({\bf r'},t) \rangle,
\label{e1}
\end{equation}     
where $N$ is the total particle number and 
\begin{equation}
\rho ({\bf r},t) = \sum_{i=1}^N \delta [{\bf r}-{\bf r}_i(t)]
\label{e2}
\end{equation}
the local density of point particles at the positions ${\bf r_i}$ at time $t$.

The density-density autocorrelation function is related to the pair 
correlation or radial distribution function,  
\begin{equation}
g({\bf r}) = \frac{1}{N}\> \langle \sum_{i\neq j}^N \delta 
({\bf r}+{\bf r_i}-{\bf r_j}) \rangle,
\label{e3}
\end{equation}
by
\begin{equation}
G({\bf r}) = g({\bf r}) + \delta ({\bf r}).
\label{e4}
\end{equation}

\begin{figure}
\centerline{\psfig{figure=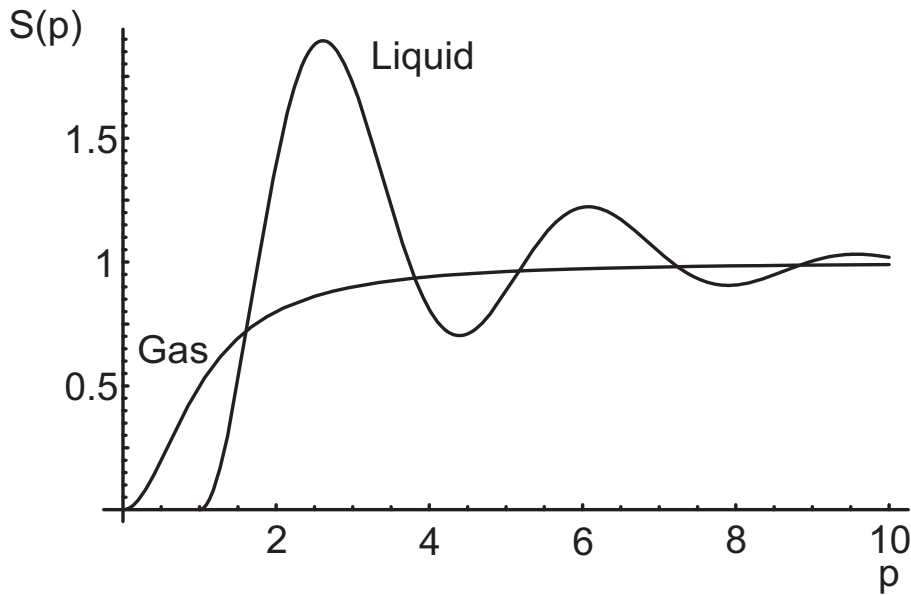,width=12cm}}
\vspace*{0.5cm}
\caption{Sketch of the static structure functions vs. momentum in the gas and liquid phase 
in arbitrary units [5].}
\end{figure}

The static structure function, defined by 
\begin{equation}
S({\bf p}) = \frac{1}{N}\> \langle \rho ({\bf p}) \rho (-{\bf p}) \rangle 
\label{e5}
\end{equation}
with the Fourier transformed particle density, 
\begin{equation}
\rho ({\bf p}) = \int d^3r \> \rho ({\bf r})\> e^{-i {\bf p} \cdot {\bf r}},
\label{e6}
\end{equation}
is the Fourier transform of the density-density autocorrelation function
\begin{equation}
S({\bf p}) = \int d^3r \> G({\bf r})\> e^{-i {\bf p} \cdot {\bf r}}.
\label{e7}
\end{equation}

For uncorrelated particles the structure function is constant for ${\bf p}\neq 0$
\cite{Ichimaru2}. The typical behavior of the static structure function in an
interacting gas and in a liquid is shown in Fig.1 according to Ref.\cite{Ichimaru}. 
There a one-component Yukawa system was assumed. 
Note that in a one-component Yukawa system there is no gas-liquid phase transition   
but only a supercritical fluid exists. 
After all the structure function shows a gas like behavior for small coupling parameters,
whereas a peak structure (fluid like behavior) develops only at rather high 
values of $\Gamma$ \cite{Ichimaru}. 
Only in the presence of repulsive and
attractive forces at the same time a gas-liquid transition shows up, ending at a critical point.
In the QGP, where a gas-liquid phase transition  
could exist \cite{Thoma3}, the appearance of peaks in the static structure function could correspond
to a real liquid phase.  

The time-dependent generalization of the density-density autocorrelation 
function, 
\begin{equation}
G({\bf r},t) = \frac{1}{N}\> \int d^3r' \> \langle \rho ({\bf r} + {\bf r'},t)
\rho ({\bf r'},0) \rangle,
\label{e8}
\end{equation}
is called van Hove function and its Fourier transform,
\begin{equation}
S({\bf p}, \omega) = \int dt \int d^3r \> G({\bf r},t)\> 
e^{i(\omega t -{\bf p}\cdot {\bf r})},
\label{e9}
\end{equation}
dynamic structure function. It is convenient to decompose the van Hove 
function into a self and a distinct part
\begin{equation}
G({\bf r},t) = G_s({\bf r},t) + G_d ({\bf r},t),
\label{e10}
\end{equation}
with
\begin{eqnarray}
G_s({\bf r},t) &=& \frac{1}{N}\> \langle \sum_{i=1}^N \delta 
[{\bf r}+{\bf r_i(0)}-{\bf r_i(t)}] \rangle,\nonumber \\
G_d({\bf r},t) &=& \frac{1}{N}\> \langle \sum_{i\neq j}^N \delta 
[{\bf r}+{\bf r_i(0)}-{\bf r_j(t)}] \rangle .
\label{e11}
\end{eqnarray}
The distinct part looks similar as the pair correlation function, i.e., showing
a pronounced peak plus one or two broad side peaks in the liquid case, if $t\ll \tau$ where
$\tau$ is the characteristic relaxation time of the system. For large
$t$, however, it becomes a smooth and flat function of ${\bf r}$.

\section{Structure functions in a quark-gluon-plasma}

Now we want to define these correlation functions in the case of the QGP. For
simplicity we consider here only the quark component. The local quark
density is given by
\begin{equation}
\rho({\bf r},t) = \bar q({\bf r},t) \gamma _0 q({\bf r},t),
\label{e12}
\end{equation}
where $q$ and $\bar q$ are the quark wave functions. In a homogeneous and
isotropic plasma, in which $\langle \rho ({\bf r})\rho ({\bf r'})\rangle $ depends only
on ${\bf r}-{\bf r'}$, the static quark density-density autocorrelation function 
is given by
\begin{equation} 
G({\bf r}) = \frac{1}{n} \> \langle \rho({\bf r},t) \rho (0,t) \rangle .
\label{e13}
\end{equation}
Here $n=N/V=\langle \rho({\bf r})\rangle $ is the average particle density in 
a homogeneous system.
At finite temperature its Fourier transform $\chi $ follows from \cite{KMT}
\begin{equation}
G({\bf r}) = \frac{1}{n}\> \lim_{\Delta \tau \rightarrow 0}\> T\> 
\sum_{n=-\infty}^{\infty} \int \frac{d^3p}{(2\pi )^3}\> e^{-i(\omega_n 
\Delta \tau -{\bf p}\cdot {\bf r})}\> \chi (\omega_n,{\bf p}), 
\label{e14}
\end{equation}
where $\omega_n=2n\pi T$ are the discrete Matsubara frequencies. 

The corresponding spectral function is given by \cite{KMT}
\begin{equation}
\sigma(\omega ,{\bf p}) = \frac{1}{\pi}\> {\rm Im} \chi (\omega ,{\bf p}),
\label{e15}
\end{equation}
where $\omega $ is a real number now. The static structure function
follows from the spectral function according to \cite{KMT}
\begin{equation}
S({\bf p}) = \frac{1}{n}\> \int_0^\infty d\omega \> \sigma (\omega ,{\bf p})\>
\coth \frac{\omega}{2T},
\label{e16}
\end{equation}
which is a consequence of the fluctuation-dissipation theorem \cite{Ichimaru2}.

The function $\chi (\omega, {\bf p})$ can be expressed by the one-loop
polarization diagram containing dressed quark propagators \cite{KMT}
\begin{equation}
\chi (\omega ,{\bf p}) = -2N_c\> T\> \sum_{n=-\infty}^\infty \int 
\frac{d^3k}{(2\pi )^3}\> {\rm Tr}[\gamma_0\, S_F(k_0,{\bf k})\, \gamma_0\,
S_F(\omega -k_0,{\bf p}-{\bf k})]
\label{e17}
\end{equation}
with the number of colors $N_c$ and $k_0=2ni\pi T$. This expression is 
related to the longitudinal part 
of the QCD polarization tensor containing only the quark loop, $\Pi_L=\Pi_{00}$, by
\begin{equation}
\chi (\omega , {\bf p})=-\frac{4N_c}{g^2}\> \Pi_L(\omega ,{\bf p}),
\label{e18}
\end{equation}
where $g$ is the QCD coupling constant.

The longitudinal part of the polarization tensor also determines the quark number
susceptibility $\chi_q (T)$, which is proportional to the ${\bf p}=0$ limit of the static
structure function (\ref{e16}), $\chi_q (T)= (n/2T)\> S(0)$ \cite{Chakraborty}. 
In the case of an isotropic 
and homogeneous plasma the polarization tensor depends only on $\omega$ and $p=|{\bf p}|$. 

According to (\ref{e9}) and (\ref{e14}) for $t\neq 0$, the dynamic structure 
function is given by
\begin{equation}
S(\omega ,{\bf p})= \frac{1}{n}\> \chi (\omega ,{\bf p}).
\label{e19}
\end{equation}

Using (\ref{e14}) to (\ref{e19}) the density-density autocorrelation 
and structure functions can be derived in the strongly coupled phase of the 
QGP by using lattice QCD and investigated for signatures of the liquid phase
as discussed above. In lattice calculations the correlation functions can
be derived directly whereas the structure functions and spectral densities
require a maximum entropy analysis \cite{Karsch3}.
For comparison, we calculate in the following the static 
and dynamic structure functions within perturbation theory, i.e., in a
weakly coupled QGP, where no pronounced structures in these functions
are expected.

\section{Hard thermal loop approach to the static structure function} 

Let us first consider the high-temperature or hard thermal loop (HTL) limit 
\cite{Weldon,Klimov,Braaten}. In this case $\chi $ is given by the high-temperature limit 
of the 1-loop polarization tensor with bare quark propagators, 
\begin{equation}
\chi^{\rm HTL}(\omega ,p) = \frac{2N_cN_f}{3}\> T^2\> \left [ 1-\frac{\omega}{2p}
\> \ln \frac{\omega +p}{\omega -p}\right ],
\label{e20}
\end{equation}
where $N_c$ is the number of colors, i.e., $N_c=3$ and $N_f$ the number
of light flavors in the QGP. For the interesting
regime of small times, $t\ll \tau$, corresponding to $\omega \gg p$ in
(\ref{e20}), $\chi $ behaves as
\begin{equation}
\chi^{\rm HTL} (\omega \gg p) = -\frac{2N_cN_f}{9}\> T^2\> \frac{p^2}{\omega ^2}
\label{e21}
\end{equation}
and the dynamic structure function (\ref{e19}) is a monotonically decreasing function of $\omega $.
In the case of a liquid this function also decreases with $\omega$ but shows a peak structure
in addition \cite{Hansen}. 

The spectral density, following from the imaginary part of the
polarization tensor according to (\ref{e15}), within the HTL limit
is given by
\begin{equation}
\sigma (\omega ,p) = \frac{N_cN_f}{3}\> T^2\> \frac{\omega }{p}\> 
\theta (p^2-\omega ^2).
\label{e22}
\end{equation}
The imaginary part of (\ref{e20}) corresponds to Landau damping.
The static structure function according to (\ref{e16}) reads
\begin{equation}
S(p) = \frac{2N_fT^3}{n}.
\label{e23}
\end{equation}
Here we have used consistently the high-temperature approximation, $\coth (\omega/2T)
=2T/\omega $, and $N_c=3$. The high-temperature limit corresponds to the classical 
limit, in which the polarization tensor, related to the dielectric function (see below),
can also be derived by the Vlasov equation \cite{Thoma}. Using the HTL polarization
function together with $\coth (\omega/2T)$ in (\ref{e16}) leads to the unphysical
result of a linearly increasing static structure function for large $p$. The 
constant structure function (\ref{e23}), on the other hand, indicates 
an uncorrelated QGP corresponding to an ideal gas -- the corresponding pair correlation
function (\ref{e4}) is equal to zero -- in the limit of infinitely
high temperature. This result also agrees with the free quark number susceptibility 
$\chi_q (T) = N_fT^2$ \cite{Chakraborty}. 
Let us note here that using the 1-loop polarization tensor beyond
the HTL limit \cite{Kapusta,Peshier2,Thoma4} in (\ref{e16}) also leads to an unphysical
result, namely a monotonically decreasing $S(p)$. This shows once more that
the naive use of perturbation theory for gauge theories at finite temperature
is inconsistent \cite{Braaten}.   

To go beyond the uncorrelated QGP, we resum the HTL polarization tensor (\ref{e18})
within a Dyson-Schwinger equation,
\begin{equation}
\Pi_L^* = \Pi_L^{\rm HTL}\> \sum_{n=0}^\infty (D_L^0\, \Pi_L^{\rm HTL})^n,
\label{e24}
\end{equation}
where $D_L^0=1/p^2$ is the bare longitudinal gluon propagator in Coulomb gauge
\cite{Thoma}. Using the relation between the longitudinal polarization tensor 
and the longitudinal dielectric function \cite{Thoma}, holding in particular for
the HTL limit,
\begin{equation}
\epsilon_L^{\rm HTL}(\omega ,p)=1-\frac{\Pi_L^{\rm HTL}(\omega, p)}{p^2},
\label{e25}
\end{equation}
the resummed polarization tensor is given by
\begin{equation}
\Pi_L^*(\omega ,p) = p^2\> \left (\frac{1}{\epsilon_L^{\rm HTL}(\omega ,p)}-1\right ).
\label{e26}
\end{equation}.

The static structure function, following from (\ref{e15}), (\ref{e16}), (\ref{e18}), and (\ref{e26})
together with the high-temperature approximation, $\coth (\omega/2T) =2T/\omega$,
using a Kramers-Kronig relation is given by \cite{Ichimaru2}
\begin{equation}
S(p) = \frac{4N_cT}{g^2}\> p^2\> \left [1-{\rm Re}
\frac{1}{\epsilon_L^{\rm HTL} (\omega =0, p)}\right ]
=\frac{2N_fT^3}{n}\> \frac{p^2}{p^2+m_D^2},
\label{e27}
\end{equation}
where $m_D^2 = -\Pi_L^{\rm HTL} (\omega =0,p)=N_fg^2T^2/6$ is the quark contribution 
to the classical Debye screening mass. The static structure function (\ref{e27})
starts at zero for $p=0$ and saturates at the uncorrelated
structure function (\ref{e23}) for large $p$. Such a structure function 
corresponds to a Yukawa system in the gas phase (see Fig.1). 
Indeed, the pair correlation function following from the Fourier transform 
of $S(p)-1$,
\begin{equation}
g(r) = -\frac{N_fT^3}{2\pi n}\> \frac{m_D^2}{r}\> e^{-m_Dr},
\label{e28}
\end{equation}
reproduces the Yukawa potential. The radial distribution function, giving the probability
of finding other quarks, in Ref.\cite{Ichimaru} is given by 
$1+g(r)$. Using (\ref{e28}) the radial distribution function becomes unphysical, i.e., 
negative, for small $r$ indicating the
failure of the Vlasov approach at small inter-particle distances \cite{Ichimaru2}.

In a next step one could use HTL-resummed quark propagators and quark-gluon vertices
in the polarization tensor as done for the quark number susceptibility \cite{Chakraborty}.
However, to look for a strongly-coupled liquid phase within the structure functions and
pair correlation functions, one has to adopt non-perturbative methods. At the classical 
level molecular dynamics, generalized to the relativistic case, could be useful
\cite{Ichimaru}. In general, of course, QCD lattice simulations would be the ultimate choice. 

\section{Summary}

In summary, we have presented definitions of the pair correlation function, the
density-density autocorrelation function, its time-dependent generalization
(van Hove function) and their Fourier transforms, the static
and dynamic structure functions in the case of a QGP. The latter
are closely related to the longitudinal part of the polarization
tensor. In liquids these correlation functions and their Fourier transforms
show a specific peak structure.
Hence they could be used to investigate the existence of a liquid
phase in a strongly coupled QGP, if calculated within non-perturbative
methods, in particular lattice QCD. 
Adopting the perturbative
high-temperature limit of the polarization tensor and its resummation corresponding
to the weak coupling limit, we have shown, as expected, that there is no 
indication of a liquid phase in the dynamic as well as static structure functions.
These results can be used as a reference for non-perturbative calculations. 

\bigskip

{\bf Acknowledgment:} I would like to thank W. Cassing, A. Peshier, F. Karsch, and W. Zajc
for helpful discussions.



\begin{thebibliography}{1}
\bibitem{QM} Proceedings of Quark Matter 2004, J. Phys. G {\bf 30}, S633 (2004).
\bibitem{Phenix} K. Adcox et al. (Phenix collaboration), Nucl. Phys. A {\bf 757}, 184 (2005).
\bibitem{Karsch1} F. Karsch, AIP Conf. Proc. {\bf 631}, 112 (2003).
\bibitem{Thoma} M.H. Thoma, in: {\it Quark-Gluon Plasma 2}, ed.: R.C. Hwa
(World Scientific, Singapore, 1995), p.51., {\it hep-ph/9503400}. 
\bibitem{Ichimaru} S. Ichimaru, Rev. Mod. Phys. {\bf 54}, 1017 (1982).
\bibitem{Thomas} H.M. Thomas, G.E. Morfill, V. Demmel, B. Feuerbacher, and
D. M\"ohlmann, Phys. Rev. Lett. {\bf 73}, 652 (1994).
\bibitem{Hamaguchi} S. Hamaguchi, R.T. Farouki, and D.H.E. Dubin, Phys. Rev.
E {\bf 56}, 4671 (1997).
\bibitem{Thoma1} M.H. Thoma, IEEE Trans. Plasma Sci. {\bf 32}, 738 (2004). 
\bibitem{Thoma2} M.H. Thoma, J. Phys. G {\bf 31}, L7 (2005); 
Erratum, J. Phys. G {\bf 31}, 539 (2005).
\bibitem{Molnar} D. Molnar and M. Gyulassy, Nucl. Phys. A {\bf 697}, 495
(2002). 
\bibitem{Shuryak} E. Shuryak, J. Phys. G {\bf 30}, S122 (2004).
\bibitem{Gyulassy} M. Gyulassy and L. McLerran, Nucl. Phys. A {\bf 750}, 30 
(2005).
\bibitem{Heinz} U. Heinz, AIP Conf. Proc. {\bf 739}, 163 (2005).
\bibitem{Cassing} W. Cassing and A. Peshier, Phys. Rev. Lett. {\bf 94}, 
172301 (2005).
\bibitem{Peshier} A. Peshier, private communication.
\bibitem{Karsch2} F. Karsch, private communication.
\bibitem{Hansen} J.-P. Hansen and I.R. McDonald, {\it Theory of Simple Liquids}
(2nd edition, Academic Press, London, 1986).
\bibitem{Ichimaru2} S. Ichimaru, {\it Basic Principles of Plasma Physics}, (Benjamin,
Reading, 1973).
\bibitem{Thoma3} M.H. Thoma, {\it hep-ph/0509154}.
\bibitem{KMT} F. Karsch, M.G. Mustafa, and M.H. Thoma, Phys. Lett. B {\bf 497},
249 (2001).
\bibitem{Chakraborty} P. Chakraborty, M.G. Mustafa, and M.H. Thoma, 
Eur. Phys. J. C {\ bf 23}, 591 (2002).
\bibitem{Karsch3} F. Karsch, S. Datta, E. Laermann, P. Petreczky, S. Stickan, 
and I. Wetzorke, Nucl. Phys. A {\bf 715}, 701 (2003).
\bibitem{Weldon} A.H. Weldon, Phys. Rev. D {\bf 26}, 1394 (1982).
\bibitem{Klimov} V.V. Klimov, Sov. Phys. JETP {\bf 55}, 199 (1982).
\bibitem{Braaten} E. Braaten and R.D. Pisarski, Nucl. Phys. B {\bf 337}, 569
(1990).
\bibitem{Kapusta} J.I. Kapusta, {\it Finite Temperature Field Theory}
(Cambridge University Press, New York, 1989).
\bibitem{Peshier2} A. Peshier, K.Schertler, and M.H. Thoma, Ann. Phys. (N.Y.)
{\ bf 266}, 162 (1998).
\bibitem{Thoma4} M.H. Thoma, S. Leupold, and U. Mosel, Eur. Phys. J. A {\bf 7},
219 (2000).

\end{thebibliography}
\end{document}